
\input harvmac.tex
\def\id{{\bf 1}}
\def\Id{{1 \kern -.40em 1 }}
\def\half{{1 \over 2}}
\def\dotimes{{ \mathop{\otimes }^{\displaystyle .} }}
\def\vitimes{{ \mathop{,}^{\displaystyle \otimes} }}
\def\complex{{ {\rm C} \kern -.5em
   {\raise .12ex \hbox{\vrule height 1.2ex width 0.02em depth 0ex}}
   \kern 0.5em }}
\def\real{{ {\rm R} \kern -0.45em \vrule height 1.44ex width 0.02em depth 0ex
   \kern 0.45em }}
\def\tto{{ \rightarrow }}
\def\Fq{{ Fun(SO_q(3)) }}
\def\Fqr{{ Fun(SO_q(3,\real)) }}
\def\Oqr{{ O_q^3(\real) }}
\def\FE{{ Fun(E_q(2)) }}
\def\Ct{{ \widetilde{C} }}
\def\Ch{{ \widehat{C} }}
\def\order#1{{ O(R^{-#1}) }}

{\nopagenumbers
\rightline{MIT-CTP-2294}
\rightline{March 1994}
\rightline{hep-th/9404007}
\vskip 1in
\centerline{\titlefont The quantum two dimensional Poincar\'e group}
\smallskip
\centerline{\titlefont from quantum group contraction}
\bigskip
\centerline{Philippe Zaugg
\foot{Partially supported by the Swiss National Science Foundation
and by funds provided by the U.S. Department of Energy (D.O.E.) under
contract \#DE-AC02-76ER03069.\hfill\break
e--mail: zaugg@pierre.mit.edu.} }
\bigskip
\centerline{\sl Center for Theoretical Physics}
\centerline{\sl Laboratory for Nuclear Science}
\centerline{\sl and Department of Physics}
\centerline{\sl Massachusetts Institute of Technology}
\centerline{\sl 77, Massachusetts Avenue}
\centerline{\sl Cambridge, MA 02139, USA}
\vskip 1.5in
\centerline{\bf Abstract}
\medskip
A new derivation of the algebra of functions on the two dimensional
Euclidean Poincar\'e group is proposed. It is based on a contraction
of the Hopf algebra $\Fq$, where the deformation parameter $q$ is sent
to one.
\vfill
\pageno=0
\eject}
\ftno=0

\newsec{Introduction}

Various quantum deformations of the Poincar\'e algebra or group are now
available, all relying on some contraction procedure,
and all being Hopf algebras.
At the algebra level, contracting the deformed universal enveloping algebra
$\CU_q(so(D+1))$ leads to the deformation of the enveloping Poincar\'e
algebra $\CU_\kappa(P_D)$
\ref\italy{E.~Celeghini, R.~Giachetti, E.~Sorace and M.~Tarlini, J. Math. Phys.
{\bf 31} (1990) 2548; {\bf 32} (1991) 1155, 1159.}
\ref\lnr{J.~Lukierski, A.~Nowicki, H.~Ruegg and V.~Tolstoy, Phys. Lett. {\bf
B264} (1991) 331.\hfill\break
J.~Lukierski, A.~Nowicki, H.~Ruegg, Phys. Lett. {\bf B293} (1992)
344.\hfill\break
J.~Lukierski and H.~Ruegg, {\sl Quantum $\kappa$--Poincar\'e in any
dimensions}, Geneva University preprint UGVA-DPT 1993/07--825.}
\ref\masla{P.~Maslanka, {\sl The $n$-dimensional $\kappa$--Poincar\'e algebra
and group}, Lodz University preprint.};
in these approaches, the original quantum group
parameter $q$ is sent to its classical value one, and a dimensionful parameter
$\kappa$ appears.
At the group level, the quantum group counterpart of the above quantum algebras
has been proposed in
\ref\zak{S.~Zakrzewski, {\sl Quantum Poincar\'e group related to
$\kappa$--Poincar\'e algebra}, Warsaw University preprint.},
through the quantization of the induced Poisson structure
on the classical Poincar\'e group. Also, a contraction of $SU_q(2)$ leads to a
different definition of the two dimensional Euclidean Poincar\'e group, where
the parameter $q$ survives the contraction
\ref\swz{P.~Schupp, P.~Watts and B.~Zumino, Lett. Math. Phys. {\bf 24} 1992,
141.}
\ref\bal{A.~Ballesteros, E.~Celeghini, R.~Giachetti, E.~Sorace and M.~Tarlini,
{\sl An $R$-matrix approach to the quantization of the Euclidean group $E(2)$},
Firenze preprint DFF 182/1/93.}.

Nevertheless, the duality between the quantum
Poincar\'e groups and the quantum Poincar\'e algebras is not always
established, contrary to what happens when the
deformations are based on a semi--simple Lie group or algebra. The lack of an
universal $R$--matrix (except in three space--time dimensions) complicates
the proof, and only in two dimensions is there a direct proof
\ref\masb{P.~Maslanka, {\sl The 2 dimensional quantum Euclidean group}, Lodz
University preprint.}.

In this letter, we propose a construction of the quantum algebra $\FE$, as
obtained in \zak, using
a contraction procedure starting directly from the Hopf algebra $\Fq$.
The parameter $q$ is sent to one as well, thereby mimicking closely
the universal enveloping algebra contraction.
Apart from increased computational task, there seems to be no
obstruction to carry
out the same program for other space--time dimensions and space--time
signatures
\ref\joe{Ph.~Zaugg, {\sl Quantum Poincar\'e group from quantum group
contraction}, in preparation.}.

In section two the necessary definitions of quantum group are quickly reviewed,
relying essentially on
\ref\rtf{N.~Reshetikhin, L.~Takhtadzhyan and L.~Faddeev, Leningrad Math. J.,
Vol. 1 (1990), 193.}.
In section three, a contraction of the classical group and its
associated Poisson structure is done, mainly
to illustrate the steps that will be used in the next section.
Section four deals with the quantum group
contraction leading to the definition of $\FE$.

\newsec{The Hopf algebra $\Fqr$}

The starting point is $\Fq$ \rtf, the algebra of functions on the quantum
group $SO_q(3)$, defined as the non--commutative $\complex$--algebra with unity
generated by the elements $T=(t_{ij})$
\foot{Our convention for indices is that $i,j,k=1,2,3$, whereas $a,b,c=2,3$}
quotiented by the relations
\eqn\rtt{
R_t T_1 T_2 = T_2 T_1 R_t ,}
where $T_1 = T \otimes I, T_2= I \otimes T$. The
$R$-matrix has the particular expression
\eqn\rt{
\eqalign{
R_t =& (e_{11}+e_{33}) \otimes e_{22} +
q(e_{11} \otimes e_{11} + e_{33} \otimes e_{33}) + q^{-1}
(e_{33} \otimes e_{11} + e_{11} \otimes e_{33}) \cr
  & +(q-q^{-1})(e_{21} \otimes e_{12} + e_{31} \otimes e_{13} +
e_{32} \otimes e_{23}) + e_{22} \otimes I \cr
  & - (q-q^{-1}) ( q^{-\half} e_{21} \otimes e_{23} +
q^{-1} e_{31} \otimes e_{13} + q^{-\half} e_{32} \otimes e_{12}) .} }
with $e_{ij}$ the basis of $3\times 3$ matrices.
That this is the orthogonal group is encoded in the supplementary
constraints
\eqn\ortho{
T C T^T C^{-1} = C T^T C^{-1} T = \id, \qquad \hbox{with} \qquad
C=\left( \matrix{0 & 0 & q^{-\half} \cr 0 & 1 & 0 \cr
q^\half & 0 & 0} \right) . }
The Hopf algebra structure is further defined by the homomorphisms
$\Delta, \epsilon$ and anti--homomorphism $S$:
$$
\Delta (T) = T \dotimes T, \qquad \epsilon(T)=\id, \qquad
S(T)= C T^T C^{-1}.
$$

The quantum three dimensional complex plane $O_q^3(\complex)$
is defined as the non--commutative $\complex$--algebra with unity
generated by the elements $x_1,x_2,x_3$,
subject to the relations
\eqn\xplane{
x_1 x_2 =  q x_2 x_1, \qquad x_2 x_3 = q x_3 x_2, \qquad
x_1 x_3 - x_3 x_1 = (q^{-\half} - q^\half) x_2^2. }
There is a natural action of the quantum group on this quantum vector
space by mean of the algebra homomorphism
$\delta : \Fq \rightarrow \Fq \otimes O_q^3(\complex)$, where
$$
\delta(x_i) = t_{ij} \otimes x_j .
$$
It is easy to see that the quadratic form $x^T C x$ in
$O_q^3(\complex)$ is preserved by this
action of the quantum group.

The compact real form of $\Fq$, for real $q$, is obtained by
endowing the algebra with the anti--involution
$$
T^* = C^T T (C^{-1})^T .
$$
Accordingly, the anti--involution $x^* = C^T x$ on
$O_q^3(\complex)$ defines the real quantum Euclidean plane
$O_q^3(\real)$. Since $\delta^*(x)=\delta(x^*)$, and the action of
$\Fqr$ preserves the quadratic form, the definition of the
two-dimensional quantum unit sphere as the quotient of $\Oqr$ by the
relation $x^{*T}x=\id$ is meaningful.

For our purpose, it will be more convenient to work in a real basis for
the quantum plane. We choose the new real generators in $\Oqr$:
\eqn\xtoz{
z_1 = {1 \over \sqrt{2}}(x_1+q^\half x_3), \qquad
z_2 = {-i \over \sqrt{2}}(x_1-q^\half x_3), \qquad z_3 = x_2 , }
or, in short, $z=M x$. Similarly, we take new generators $V=M T M^{-1}$
in $\Fqr$, now real $V^*=V$, and define $R_v = (M \otimes M) R_t
(M^{-1} \otimes M^{-1})$, so that the constraints \rtt\
become $R_v V_1 V_2 =
V_2 V_1 R_v$, and $\delta(z_i) = v_{ij} \otimes z_j$. Only the
orthogonality relations \ortho\ are significantly affected
\eqn\orthoz{
\eqalign{
V \Ch V^T &= \Ch, \qquad\quad \hbox{with} \qquad \Ch=M C M^T \cr
V^T \Ct V &= \Ct, \qquad\quad \hbox{with} \qquad \Ct=M^{-1T} C M^{-1} ,} }
as well as the antipode $S(V)= \Ch V^T \Ct$, and the quantum plane relations
\xplane
\eqn\zplane{
z_3 z_2 - q z_2 z_3 = i (q z_1 z_3 - z_3 z_1) , \qquad
z_1 z_2 - z_2 z_1 = i(1-q)z_3^2 .}
Note that the second relation in \xplane\ is nothing but the *-conjugate
of the first, so we skip it here.

As $\Fqr$ is a non--commutative deformation of the algebra
of functions on the classical Lie group $SO(3,\real)$, the $R$--matrix
$R_v$ induces a Poisson structure on that latter algebra
through
\eqn\pbr{
\{ \Phi \vitimes \Phi \} = [ \Phi \otimes \Phi, r ], }
where the classical $r$--matrix is defined by $r = h^{-1}(R_v-1)
\hbox{~mod~} h$, $\Phi =(\phi_{ij})$ are the matrix elements
of the classical group and $h$ is a parameter governing the deformation,
$h \rightarrow 0$ being the classical limit
\ref\dri{V.~Drinfeld, {\sl Quantum groups},
ICM Proceedings vol 1, Berkeley 1986.}.

Having in mind the contraction procedure of the next sections,
we set the parameter
$q=\exp(\gamma / R)$, as for the universal envelopping algebra
contraction, with $R$ the radius of the sphere (or de Sitter radius), and take
$i\gamma$ as our deformation parameter. Since we are dealing with a real
algebra, the $i$ factor is necessary if we
follow the definition of algebras deformations in \dri\ and require
$\{a,b\}^* = \{a^*,b^*\}$.

{}From the expression of $R_v$, it is not difficult to extract the
classical $r$--matrix
\eqn\rcl{
r = {1 \over R} \left( X_1 \otimes X_2 -  X_2 \otimes X_1 +
i X_k \otimes X_k \right)  =
{1 \over R} X_1 \wedge X_2 + (\dots) }
in the conventional basis for the Lie algebra $so(3,\real)$ for the three
dimensional representation, $(X_i)_{jk}
= \epsilon_{ijk}$.
The last term in \rcl\ does not contribute to the Poisson bracket \pbr\
since it is $SO(3,\real)$ ad--invariant.

\newsec{Classical group contraction}

Now we perform a $SO(3,\real)$ contraction in order to define a Poisson
structure on the two dimensional Euclidean Poincar\'e group. An element of
$SO(3,\real)$ will be parametrized by the three rotation angles around the
respective $z_i$ axis, $\Phi(\theta_k)=\exp(\sum_1^3 \theta_k X_k)$
\ref\gil{R.~Gilmore, {\sl Lie groups, Lie algebras and some of their
applications}, New-York, Wiley, 1974.}.
Geometrically,
one considers the action of $SO(3,\real)$ on the
two--sphere of radius $R$ in the vicinity of the point $p=(R,0,0)$, and
let the radius $R \rightarrow \infty$. Simultaneously, one imposes that
the rotation angles $\theta_{2,3}$ go to zero like $1/R$,
so that the transformation of a point in the vicinity of $p$
remains at a finite distance from $p$. We therefore define
$(\theta_k)=(\theta,\alpha/R,\beta/R)$.

{}From the explicit expression of $\Phi$ \gil, the matrix elements $\phi_{ij}$
have the expansion
\eqn\expphi{
\phi_{ij}(\theta_k) = \sum_{n=0}^\infty {\phi_{ij}^n(\theta,\alpha,\beta)
\over R^n} ,}
where, depending on the indices $i,j$, the series contains either odd or even
powers of $R$.
Then, in the limit of infinite radius,
$$
\eqalign{
  \left( \matrix{1 \cr z_2' \cr z_3'} \right) & =
\lim_{R \rightarrow \infty} \left( \matrix{z_1'/R \cr z_2' \cr z_3'} \right)=
\left( \matrix{ 1 & 0 & 0 \cr \phi_{21}^1 & \phi_{22}^0 & \phi_{23}^0 \cr
\phi_{31}^1 & \phi_{32}^0 & \phi_{33}^0 } \right)
\left( \matrix{1 \cr z_2 \cr z_3} \right) \cr
  & = \left( \matrix{ 1 & 0 & 0 \cr -\beta s/\theta +
\alpha(1-c)/\theta & c &
s \cr \alpha s/\theta + \beta(1-c/\theta & -s & c} \right)
\left( \matrix{1 \cr z_2 \cr z_3} \right) }
$$
where $s=\sin\theta, c=\cos\theta$. Apart from the akward parametrization of
the translations, this
is the usual representation for the classical two dimensional Euclidean
Poincar\'e group, $z_{2,3}$ being the coordinates of the two dimensional
space--time.

In order to fully specify the Poisson structure on the Poincar\'e group, we
only need
the Poisson brackets for the matrix elements $\phi_{ij}^n$
entering the above equation.
This is
achieved by plugging the expansion \expphi\ in the $SO(3,\real)$ Poisson
brackets \pbr\ and matching terms of equal power in $R$. Dropping the
superscripts and setting $\phi_a=\phi_{a1}$, we get for the non--vanishing
Poisson brackets
\eqn\pbrE{
\eqalign{
\{ \phi_{22} , \phi_{2} \} &= -(\phi_{23})^2 \cr
\{ \phi_{23} , \phi_{2} \} &= \phi_{22} \phi_{23} \cr
\{ \phi_{32} , \phi_{2} \} &= -\phi_{23} \phi_{33} \cr
\{ \phi_{33} , \phi_{2} \} &= \phi_{23} \phi_{32} \cr
\{ \phi_{2~} , \phi_{3} \} &= -\phi_{3} }
\qquad\qquad
\eqalign{
\{ \phi_{22} , \phi_{3} \} &= -\phi_{32} -\phi_{23}\phi_{33} \cr
\{ \phi_{23} , \phi_{3} \} &= -\phi_{33} +\phi_{22}\phi_{33} \cr
\{ \phi_{32} , \phi_{3} \} &= \phi_{22}-(\phi_{33})^2 \cr
\{ \phi_{33} , \phi_{3} \} &= \phi_{23}+\phi_{32} \phi_{33} \cr
\vphantom{\phi_{2}} } }
Using the orthogonality relations (in the limit $R \tto \infty$), this can be
neatly recast in the compact form
\eqn\pbrEc{
\eqalign{
  \{ \phi_{ab},\phi_{cd} \} & = 0 \cr
  \{ \phi_{ab},\phi_{c~} \} &= \left( (\phi_{a2}-\delta_{a2}) \phi_{cb} +
\delta_{ac}(\phi_{2b}-\delta_{2b}) \right) \cr
  \{ \phi_{2~}, \phi_{3~} \} &= - \phi_3 } }
This is similar to
the expression obtained in \zak, with some obvious relabelling.

\newsec{Quantum group contraction}

Turning to the quantum case, we will proceed in much the same way as in the
previous section. We consider elements of $\Oqr$ living on the two dimensional
quantum sphere
\eqn\metric{
z^T \Ct z = {1+q^{-1} \over 2}( z_1^2 + z_2^2) + {q+q^{-1} \over 2} z_3^2
= \CR^2 .}
It is convenient to absorb an irrelevant factor in $R^2 = 2\CR^2/(1+q^{-1})$.
As mentioned earlier, the contraction also involves taking
$q$ simultaneously to its classical value by letting $q=\exp(\gamma/ R)$.
Here again, we consider the vicinity of the point $(z_i)=(R,0,0)$, $R \tto
\infty$, and \metric\ allows us to expand $z_1$ as a series in $R$
\eqn\zexp{
z_1 = R \left(\id- {1 \over 2R^2}(z_2^2+z_3^2) + \order{3} \right) ,}
where $z_2,z_3$ are finite ({\it i.e.} of order 1).
Inserting this expansion
in \zplane, we observe that the divergent terms cancel, and the finite part of
these expressions yields the constraint
\eqn\Eplane{
[ z_2, z_3 ] = -i \gamma z_3 .}
It is worth pointing out that, in contrast to the classical case, the choice
of $z_1$ as the diverging coordinate is not arbitrary. This is due
to the non--commutativity of the $z$, which for instance links $[z_1,z_2]$ to
$z_3$ \foot{Taking $z_3 \tto \infty$ with $z_{1,2}$ finite can not be
compensated by our behaviour of $q$ alone, and would require a singular
change of variables in \xtoz, which would in turn cause troubles in the
contraction of $RVV=VVR$.}.

Inspired by \expphi, we rewrite the generators of $\Fqr$ as an expansion in
the contraction parameter $R$
\eqn\vexp{
v_{ij} = \sum_{n=0}^\infty {v_{ij}^n \over R^n} .}
{}From simple requirements we will collect enough informations on the
$v_{ij}^n$ to enable us to derive all the necessary relations characterizing
the algebra $\FE$.
First, we require that under the action of $\Fqr$ by
mean of the mapping
$\delta$, the elements $z_{2,3}$ remain of order 1 in the limit
$R \tto \infty$. Since
$$
\delta(z_a) = v_{a1} \otimes z_1 + v_{ab} \otimes z_b ,
$$
this is only possible if $v_{a1}^0 = 0$. Next we apply $\delta$ on both sides
of \zexp\ to get
\eqn\deltaz{
{1 \over R}(v_{11} \otimes z_1 + v_{1a} \otimes z_a ) = \Id \otimes \id -
{1\over R^2}(\delta(z_2)^2+\delta(z_3)^2) + \order{3}, }
which implies that $v_{11}^0 = \Id$ and $v_{11}^1 \otimes \id + v_{1a}^0
\otimes z_a = 0$, since $\delta(z_{2,3})$ are finite.
Feeding the first orthogonality relations \orthoz\
with this partial knowledge allows us conclude, by considering the order 1
term, that $v_{1a}^0=0$, and thus $v_{11}^1=0$, with the help of \deltaz.

Gathering all this, we can take the $R \tto \infty$ limit in $\delta(z) = V
\mathop{\otimes}\limits^{.} z$, provided we divide $z_1$ by $R$, which yields
\eqn\Etrans{
\delta \left( \matrix{\id \cr z_2 \cr z_3} \right) =
\left( \matrix{\Id & 0 & 0 \cr v_{21}^1 & v_{22}^0 & v_{23}^0 \cr
v_{31}^1 & v_{32}^0 & v_{33}^0 } \right) \dotimes
\left( \matrix{\id \cr z_2 \cr z_3} \right) .}
It is then natural to take the elements $\Id, u_{ab} = v_{ab}^0$ and $u_a =
v_{a1}^1$ as the generators of $\FE$, the algebra of functions on the quantum
Euclidean group $E_q(2)$.

We still have to take into account the constraints
imposed by the relation $R_v V_1 V_2 = V_2 V_1 R_v$. Expanding the
$R_v$--matrix in $R$ and matching the appropriate powers of $R$, we get from
the order 1 term that $v_{ab}^0$ commute among themselves. From the order
$1/R$ term and the orthogonality relations at order $1$, we get relations
similar to \pbrE, and from the $1/R^2$ term together with orthogonality at
order 1 and $1/R$, we get $[v_{31}^1,v_{21}^1]= i \gamma v_{31}^1$.
All this is summarized in the compact form
\eqn\defE{
\eqalign{
  [ u_{ab}, u_{cd} ] &= 0 \cr
  [ u_{ab}, u_{c~} ] &= i \gamma \left( (u_{a2}-\delta_{a2}) u_{cb} +
\delta_{ac}(u_{2b}-\delta_{2b}) \right) \cr
  [ u_{2~}, u_{3~} ] &= -i \gamma u_3 } }
These are all the relations containing only the elements $v_{ij}^{0,1}$. There
are of course many other relations involving higher order elements $v_{ij}^n$,
but these do not add new constraints.

The limit of the orthogonality relations \orthoz\ amount to say that the
matrix $U=(u_{ab})$ is an ordinary orthogonal matrix
\eqn\orthoE{
U^T U = \Id = U U^T .}
The algebra $\FE$ has a Hopf algebra structure inherited from its
pre--contracted ancestor $\Fqr$, which reads
\eqn\hopfE{
\eqalign{
  \Delta(U) &= U \dotimes U \vphantom{U^T} \cr
  \Delta(u) &= u \dotimes \Id + U \dotimes u \vphantom{U^T} }
\quad\qquad
\eqalign{
  \epsilon(U) &= \Id \vphantom{U^T} \cr
  \epsilon(u) &= 0 \vphantom{U^T} }
\quad\qquad
\eqalign{
  S(U) &= U^T \vphantom{U^T} \cr
  S(u) &= - U^T u \vphantom{U^T} } }
where $u$ is the column vector $(u_2 u_3)^T$. Note that in deriving these
expressions, one frequently encounters elements which are not part of the
$\FE$ algebra as defined in \Etrans, like for instance
$$
S(u_2) = S(v_{21}^1) = v_{12}^1 - {i \over 2} \gamma(v_{22}^0-\Id).
$$
But with the help of the orthogonality relations, we can always
trade these foreign elements for elements belonging to $\FE$.

We conclude that the algebra $\FE$ defined by the equations
\defE,\orthoE\ and \hopfE\ and equipped with the anti--involution
$$
U^*=U \qquad\qquad u^* = u,
$$
is a *--Hopf algebra. In addition, the mapping $\delta$ in \Etrans\ defines
an action of $\FE$ on the quantum real Poincar\'e plane generated by the
elements $z_{2,3} = z_{2,3}^*$ subject to the constraint \Eplane.
This algebra is the same as in \zak, but here we derived it only using a
contraction on $\Fqr$. It differs from \zak\ in the absence of $\gamma$ in
\pbrEc, the presentation here being consistent with the definition of
algebras deformations in \dri.

\newsec{Conclusion}

This alternative way of obtaining $\FE$ shed light on why the direct
quantization of \zak\ works and is linear in $\gamma$: here this parameter
appears only in the ratio $\gamma/R$, and the contraction amounts essentially
to an expansion in $R$, keeping only the
lowest order terms. Therefore, no higher powers of $\gamma$ arise.
Since this approach is very similar in spirit to the method of \lnr\
--~we start from a dual structure and take the same limit for $q$~--
we believe that it yields the quantum group dual to the
$\kappa$--Poincar\'e algebra, once generalized for higher dimensions.
At the moment, this is true for the two dimensional case.

\medskip\noindent
{\bf Acknowledgments.} I am indebted to R.~Jackiw and M.~Bergeron for many
useful discussions.
\listrefs
\bye